\documentclass[prd,aps,showpacs,tightenlines,nofootinbib,preprint,preprintnumbers,amsmath,]{revtex4}
\usepackage{graphicx}
\usepackage{amsfonts}
\usepackage{amssymb}
\usepackage{latexsym}
\usepackage{cancel}
\usepackage{color}
\input{colordvi.tex}

\newcommand{\bear}{\begin{array}}  \newcommand{\eear}{\end{array}}
\newcommand{\bea}{\begin{eqnarray}}  \newcommand{\eea}{\end{eqnarray}}
\newcommand{\beq}{\begin{equation}}  \newcommand{\eeq}{\end{equation}}
\newcommand{\bef}{\begin{figure}}  \newcommand{\eef}{\end{figure}}
\newcommand{\bec}{\begin{center}}  \newcommand{\eec}{\end{center}}

\newcommand{\bib}{\bibitem}

\newcommand{\siml}{\lesssim}

\def\IBB#1#2#3{{\bf #1}, #2 (20#3)}

\def\APP#1#2#3{Ann. Phys. {\bf #1}, #2 (20#3)}
\def\APJ#1#2#3{Astrophys. J. {\bf #1}, #2 (19#3)}

\def\APJSS#1#2#3{Astrophys. J. Suppl. {\bf #1}, L#2 (20#3)}

\def\CQGG#1#2#3{Class. Quantum Grav. {\bf #1}, #2 (20#3)}

\def\MPLA#1#2#3{Mod. Phys. Lett. A {\bf #1}, #2 (19#3)}

\def\PRD#1#2#3{Phys. Rev. D {\bf #1}, #2 (19#3)}
\def\PRDD#1#2#3{Phys. Rev. D {\bf #1}, #2 (20#3)}

\def\PRLL#1#2#3{Phys. Rev. Lett. {\bf#1}, #2 (20#3)}

\def\PRTT#1#2#3{Phys. Rep. {\bf#1}, #2 (20#3)}

\definecolor{mod1}{rgb}{1,0,0}
\definecolor{mod2}{rgb}{0,0,1}
\definecolor{mod3}{rgb}{0,.5,0}
\definecolor{modAT}{rgb}{0.5,0,0.5}

\begin{document}


\title{Effective Search Templates for 
a Primordial Stochastic Gravitational Wave 
Background}
\author{Takeshi Chiba}
\affiliation{Department of Physics, College of Humanities and Sciences,
Nihon University, Tokyo 156-8550, Japan}
\author{Yoshiaki Himemoto}
\affiliation{Center for Educational Assistance, 
\\ Shibaura Institute of Technology, Saitama 337-8570, Japan}
\author{Masahide Yamaguchi}
\affiliation{Department of Physics and Mathematics, Aoyama Gakuin
University, Sagamihara 229-8558, Japan}
\author{Jun'ichi Yokoyama} 
\affiliation{Research Center for the Early Universe (RESCEU), Graduate
School of Science, The University of Tokyo, Tokyo 113-0033, Japan}

\date{\today}
\preprint{\cr RESCEU-10/07 }
%

\begin{abstract}
We calculate the signal-to-noise ratio (SNR) of 
the stochastic gravitational-wave background in an extreme case that 
its spectrum has a sharp falloff with its amplitude close to the 
detection threshold. Such a spectral feature is a characteristic 
imprint of the change in the number of relativistic degrees of 
freedom on the stochastic background generated during inflation 
in the early Universe. We find that, although SNR is maximal with 
the correct template which is proportional to the assumed real spectrum, 
its sensitivity to the shape of template is fairly weak indicating that 
a simple power-law template is sufficient to detect the signature.
\end{abstract}

\pacs{98.80.Cq, 04.30.-w, 04.80.Nn}

\maketitle


\section{Introduction}
The detection of a stochastic background of primordial gravitational
waves (GWs) is an exciting challenge. Since the GWs are decoupled from
other ingredients of the Universe
 after the Planck time, their detection enables us to
probe the early Universe long before the recombination, providing
much information of the early Universe and 
high energy physics at the epoch when they are generated \cite{GW}.  
For example, 
we can probe the equation of  state $w=p/\rho$ in the early Universe
using the spectrum of a stochastic GW background \cite{SY}, and
lepton asymmetry can be evaluated by investigating effects 
of neutrino free streaming on the power spectrum of primordial GWs \cite{IYY}.

One of the most probable sources of such primordial GWs is inflation.
Inflation was proposed as the most natural solution to the
difficulties of the standard big bang cosmology, such as the horizon
problem and the flatness problem \cite{inflation}. It also generates
primordial gravitational waves (tensor modes) \cite{staro} 
as well as the
primordial density fluctuations (scalar modes) \cite{pert}. The
latter has been observed as the cosmic microwave background
anisotropies by the cosmic background explorer (COBE) satellite
\cite{COBE}, the Wilkinson microwave anisotropy probe (WMAP) satellite
\cite{WMAP,WMAP3}, and so on. 
On the other hand, the
tensor fluctuations have not yet been detected and only the upper
limit on the ratio of the amplitude of tensor fluctuations to scalar
fluctuations is obtained as $r < 0.55$ (95\% C.L.) at $k = 0.002 {\rm
  Mpc}^{-1}$ \cite{WMAP3}.

The detection of tensor fluctuations (GWs) generated during inflation
is very important in that their amplitudes can determine the energy
scale of inflation directly, while scalar fluctuations are sensitive
to a combination of the potential energy and its derivative \cite{LL}.
The energy scale of the inflation strongly constrains inflation models
proposed so far. Furthermore, the consistency relation \cite{LL} which
relates the ratio of the amplitude of tensor fluctuations to scalar
fluctuations $r$ to the spectral index of tensor fluctuations $n_{T}$,
if confirmed, would be an extremely important signature of
single-field inflation and could help to discriminate  from
other mechanisms for the generation of the spectra such as a curvaton
mechanism \cite{curvaton} or
a cyclic universe \cite{cyclic}.

The observational programs for detecting such GWs have been proposed
as the next generation projects. For example, DECIGO is proposed in
Japan \cite{DECIGO} and the big bang observer (BBO) at NASA
\cite{BBO}.  Since the typical
amplitude of a stochastic background of primordial GWs is extremely
small, the signal to noise ratio (SNR) is not so large even if noises
are suppressed below the quantum level (ultimate DECIGO). Thus, in
order to enhance the SNR, search templates as well as
cross-correlation analysis are necessary for observations of a
stochastic GW background \cite{AR}.

Although the spectrum of GWs generated during inflation may well be
approximated by power-law shape, the present-day spectrum is 
different from
the original one.  The amplitude of GWs is redshifted by the cosmic
expansion and the expansion rate depends on the matter content of the
Universe. The spectrum is roughly proportional to $f^{-2}$, $f^0$,
$f$ for the modes which reenter the horizon in the matter, 
the radiation, and the kinetic-energy dominated
phases, respectively, 
where $f$ is the frequency \cite{GW,TCS}. Therefore, templates
with (broken) power-law shapes are considered and their improvements of
SNR have been discussed \cite{Bose}.  In reality, however, the
present-day spectrum is {\it never} power law nor smooth 
even if the initial spectrum is power law.  
The effective number of relativistic degrees of freedom changes with
temperature and these changes leave characteristic features in the
spectrum \cite{WK}.\footnote{The effect of quark gluon plasma phase
  transition is discussed in \cite{QGP}.} The changes of the effective
number of degrees of freedom depending on the mass thresholds induce
relatively rapid
 changes of the amplitude of the spectrum, rather than the
change of the power-law index.  
Therefore, it is important to assess the
validity of the use of the templates of power-law shape as search
templates for a stochastic gravitational-wave background.

In this paper, we consider the frequency ranges of
$10^{-4}-10$Hz suitable for proposed future detectors of GWs
(LISA,DECIGO,BBO). These frequency ranges correspond to the modes
which reentered into the horizon at the temperature 
$100{\rm GeV}-10^4$TeV and
the spectrum is damped due to the electroweak phase transition.
In the next section, we briefly review primordial GWs produced during
inflation and search templates to detect them. In Sec. III, we compare
SNRs for templates with a simple power-law type to those for templates
with a rapid change of the amplitude and discuss how effective the former
templates are for not only detection of such GWs but also probing the
change of the amplitude of their spectrum. We give discussions and
summary in the final section.

\section{Detecting a stochastic background of gravitational waves}

\subsection{Property of a stochastic background of GWs produced during
  inflation}

In order to discuss the property of an isotropic and homogeneous
stochastic background of GWs, one introduces the dimensionless
quantity $\Omega_{\rm gw}(f)$ that is the energy density of a GW 
stored in a logarithmic frequency interval around $f$
divided by
the critical energy density. In terms of the characteristic amplitude
of a GW, $h_{\rm c}$, this quantity is
expressed as
\begin{equation}
 \Omega_{\rm gw}(f) 
=\frac{2\pi^{2}}{3H_{0}^{2}}f^{2}|h_{\rm c}|^{2} \,,
\label{eq:omega}
\end{equation}
where 
$H_{0}\cong 72{\rm km/sec/Mpc}=2.3\times 10^{-18}{\rm /sec}$ is the present Hubble parameter. 
The spectrum of stochastic background of GWs generated quantum
mechanically during inflation is calculated by the quantum field
theory of a massless minimally coupled field in inflationary
background \cite{staro}. 
However, we obtain some insights into the spectrum shape
without  detailed calculation. 
Until a mode reenters the Hubble radius, the characteristic amplitude
$h_{\rm c}$ takes a constant value proportional to the Hubble parameter,
 $H_{\rm inf}$, when the mode left the Hubble radius during
inflation.  On the other hand, the amplitude damps as 
 $1/a$ after the mode reenters the Hubble radius.
 Therefore, the present characteristic
amplitude is given by
\begin{equation}
|h_{c}| \simeq \sqrt{\frac{8}{\pi}}\frac{H_{\rm inf}}{M_{Pl}}
 \frac{a(t_{k})}{a(t_{0})} .  
\label{eq:c_amplitude}
\end{equation}
Here, $t_{0}$ is the present 
time and $t_k$ is the epoch when the mode (with wave number $k$) 
reentered the Hubble radius, $2 \pi f = k = a(t_{k}) H(t_{k})$. 
If at $t_k$ the Universe is dominated by an ingredient
 with the equation of state $w$, then 
(naively)  from the Friedmann equation we obtain $H(t_k)\propto a(t_k)^{-3(1+w)/2}$ 
and thus $f\propto a(t_k)^{-(1+3w)/2}$.  
Therefore the shape of the present density parameter $\Omega_{\rm gw}(f)$ is given by
\bea
\Omega_{\rm gw}(f) \propto a(t_k)^{1-3w}\propto f^{-2(1-3w)/(1+3w)}.
\eea
Thus we recover 
$\Omega_{\rm gw}\propto f^0,f^{-2},f^1$ for $w=1/3,0,1$, respectively. 

This naive estimate neglects the effect of the change of relativistic degrees of freedom 
during the radiation dominated epoch \cite{WK}. During the radiation dominated epoch, 
the energy density of relativistic particle $\rho_{\rm rad}=(\pi^2/30)g_*T^4$ 
does not scale as $a^{-4}$. {}From the entropy conservation \cite{kt}
\bea
\frac{\pi^2}{45}g_{*S}T^3a^3={\rm constant},
\eea
we obtain
\bea
\rho_{\rm rad} \propto g_*g_{*S}^{-4/3}a^{-4}.
\eea
Here $g_*$ and $g_{*S}$ account for the total number of effective relativistic degrees of freedom. 
As long as the Universe is fully thermalized, these two coincide with
each other. 
Taking account of the effective relativistic degrees of freedom, for the mode entering into 
the horizon during the radiation era, we have
\bea
\Omega_{\rm gw}\propto f^2 a(t_k)^2\propto a(t_k)^4H(t_k)^2\propto g_*(t_k)g_{*S}(t_k)^{-4/3}.
\label{omegagw}
\eea
Since $g_*$ and $g_{*S}$ coincide for $100{\rm GeV} <T<10^3{\rm TeV}$, 
we find $\Omega_{\rm gw}\propto g_{*}(t_k)^{-1/3}$. 

\subsection{Detection method of a stochastic background of GWs}

In general, the GW background signal is expected to be very week and
is usually masked by the detector noises.  To detect such tiny
signals, it is practically impossible to detect the signal from the
single-detector measurement.  Thus, we cross correlate the two outputs
obtained from the different detectors and seek a common signal.  We
denote the detector outputs by $s_{i}$ with
\begin{equation}
s_{i}(t)=h_{i}(t)+n_{i}(t)\,,  
\label{output}
\end{equation}
where $i=1,2$ corresponds to the $i$-th detector, and $h_{i}(t)$ is the
gravitational-wave signal and $n_{i}(t)$ is the noise.  Then, the
cross-correlation signal $S$ is given by multiplying the outputs of
the two detectors and integrating over the observational time:
\begin{eqnarray}
    S
&\equiv&
    \int^{T/2}_{-T/2} dt  \int^{T/2}_{-T/2} dt'  
    s_{1}(t) 
        s_{2}(t') Q(t-t')\,, 
\label{eq:correlation}
\end{eqnarray} 
where the filter function $Q$ is introduced to enhance the
detectability of the GW signals and we take the observation time $T$
to be large enough. We also assume that the statistical property of both 
the GW signal and the noise is stationary, which implies that the argument
of $Q$ depends only on the time difference $t-t'$.

The detectability in the context of stochastic background searches is
quantified by the signal-to-noise ratio for the cross-correlation
signal $S$
\begin{equation}
 {\rm SNR} =
\frac{\langle S \rangle}
{\sqrt{\langle S^{2} \rangle-\langle S \rangle^{2}}}.
\label{eq:snr}
\end{equation}
Under the assumption that the two different detectors (or output data
stream) have no correlation of noise in the weak limit ($h_i\ll n_i$), 
the mean and variance are given by
\begin{eqnarray}
\langle S \rangle &=& 
\frac{3H_{0}^{2}}{20\pi^{2}} T \int_{-\infty}^{\infty}
df \lvert f \rvert^{-3} \Omega_{\rm gw}(\lvert f \rvert) 
\gamma(\lvert f \rvert)\tilde{Q}(f)\,,
\label{eq:mu} \\
\langle S^{2} \rangle-\langle S \rangle^{2}
 &\simeq& \frac{T}{4}\int_{-\infty}^{\infty} df
P_{1}(\lvert f \rvert)P_{2}(\lvert f \rvert)
\lvert \tilde{Q}(f) \rvert^{2} \label{eq:sigma}\,,
\end{eqnarray}
where $ \tilde{Q}(f)$ is the Fourier transform of the filter
function $Q(t-t')$ in the frequency domain. $P_{i}(\lvert f \rvert)$
is the noise power spectrum of $i$-th detector defined by 
\bea
\langle n_i(t)n_i(t')\rangle =\frac{1}{2}\int_{-\infty}^{\infty}df 
e^{2\pi if(t-t')}P_i(\lvert f \rvert).
\eea  
$\gamma(|f|)$ is the overlap reduction function which
characterizes the reduction in sensitivity to a stochastic background 
arising from the separation time delay and relative orientation of 
the two detectors \cite{ChF}. If their
orientations are coincident and coaligned without any systematic noise
correlation between them, the overlap reduction function $\gamma(|f|)$
becomes constant for all frequencies $f$. When the arms of
a detector are separated by 90 degrees,  $\gamma(|f|)=1$.

We would like to determine the functional form of $\widetilde{Q}(f)$
which maximizes $\mathrm{SNR}$. In order to find such
a function, we introduce an inner product $(A|B)$
for any pair of complex functions $A(f)$ and $B(f)$ with weight
functions $P_{1}(|f|)\cdot P_{2}(|f|)$ \cite{AR}:
\begin{equation}
(A|B)\equiv \int_{-\infty}^{\infty}\,df\,A^*(f) \, B(f)\,
\,P_{1}(|f|)\,P_{2}(|f|)\,.
\label{eq:inner_product1}
\end{equation}
As long as  $P_{i}(|f|)$ is positive for all frequencies, $(A|B)$
satisfies the same properties of ordinary inner product in
three-dimensional Euclidean space. Using Eqs.(\ref{eq:mu}) and (\ref{eq:sigma}) and 
the inner product defined in Eq.(\ref{eq:inner_product1}), 
the SNR (\ref{eq:snr}) is rewritten as \cite{AR}
\begin{equation}
\mathrm{SNR}^2 
\simeq  {\left(\frac{3H_{0}^{2}}{10\pi^{2}}\right)^{2}}
\frac{T}{{\displaystyle (\widetilde{Q}|\widetilde{Q})}}
\left(
    \widetilde{Q}\,\left| 
    \frac{\gamma(|f|)\,\Omega_{\rm gw}(|f|)}{|f|^{3}P_{1}(|f|)
    P_{2}(|f|)}\right. 
\right)^2\,.
\label{eq:SNR^2_weak}
\end{equation}
Then, regarding functions $\widetilde{Q}(f)$ and
${\gamma(|f|)\,\Omega_{\rm gw}(|f|)}/{|f|^{3}P_{1}(f) P_{2}(|f|)}$ as
three-dimensional vectors, we find that the filter function maximizing
the SNR is given by
\begin{equation}
\widetilde{Q}(f) =
\frac{\gamma(|f|)\,\Omega_{\rm gw}(|f|)}{|f|^{3}P_{1}(|f|)
    P_{2}(|f|)}\,.
\label{eq:filter1}
\end{equation}
Here we have neglected an overall normalization constant because the
SNR (\ref{eq:snr}) is independent of it. Note that $\widetilde{Q}(f)$
also becomes an even function of a frequency $f$. Hence, in the following 
we assume that $f$ is positive definite.

It should be noted that the resultant filter function
$\widetilde{Q}(f)$ depends on not only {\it known} functions such as
the overlap reduction function and the noise spectrum but also an {\it
  unknown} function, that is, the spectrum of GWs $\Omega_{\rm
  gw}(f)$. Then, it is not until we assume the spectrum of GWs that we
can determine the maximized SNR and the corresponding filter function,
which requires us to arrange search templates of the spectrum of GWs.
However, since the number of templates is limited in practice, such a
template does not necessarily coincide with the real spectrum of the GWs
background. Thus, it is very important to estimate SNRs for such
templates when the real spectrum of the GWs background is given. More
concretely, as explained in the Introduction, the spectrum of GWs
generated during inflation is in fact neither power law nor smooth.
Instead, the amplitude of the spectrum changes rapidly due to the
changes of the effective number of degrees of freedom depending on the
mass thresholds. Then, we wonder how much it is justified to use
templates with a simple power-law type, which have been often
considered. If such simple templates have  large enough SNRs, the
number of templates is significantly reduced. On the contrary, in case
their SNRs are small, we would be able to find a sudden change of
the amplitude of the spectrum if we use templates which take a sudden
change of the amplitude into account, albeit the number of the
templates becomes much larger. Therefore, we compare the SNRs of
templates with a simple power-law type to those of templates with a
realistic form of the spectrum.

\section{Effectiveness of search templates}

In this section, we compare SNRs for templates with a simple power-law
type to those for templates with the change of the amplitude for the
LIGO II and the next-generation space interferometers such as
FP-DECIGO and LISA. Then, we discuss how effective the former templates are for
not only detection of such GWs but also probing the change of the
amplitude of their spectrum.

As the spectrum of GW
background, taking into account the change of the effective number of
relativistic 
degrees of freedom, we adopt the following simple model in which 
 $\Omega_{\rm gw}(f)$ is characterized by the step
function:\footnote{In reality, the change of the spectrum is much milder than 
the step function \cite{WK}. Here we consider the extreme case 
so that the SNR might change most drastically.}
\begin{equation}
\Omega_{\rm gw}(f) =
\Omega_N
\left[
\Theta\left(1-\frac{f}{f_c}\right)
+ d \cdot \Theta\left(\frac{f}{f_c}-1\right)\right]\,,
\label{real}
\end{equation}
where $f_c$ is the critical frequency corresponding to the electroweak 
phase transition ($\sim 10^{-4}$ Hz) and $d$ is the damping factor
that reflects the change of the effective number of relativistic
degrees of freedom $g_*$, and $\Omega_N$ is a constant. 
Since $\Omega_{\rm gw}$ is proportional to
$ g_*^{-1/3}$ (Eq.(\ref{omegagw})), the damping factor due to 
electroweak phase transition in the standard model of elementary particles 
$d_{\rm SM}$ is  $d_{\rm SM}\simeq [g_*(>{\rm TeV})/g_*(1{\rm GeV})]^{-1/3}\simeq 0.9$. 
Even in the supersymmetric extension of the standard model, only $g_*(>{\rm TeV})$ 
is doubled and the damping factor becomes $d\simeq 2^{-1/3}d_{\rm SM}\simeq 0.7$. 

As mentioned in the previous section, the optimal filter is determined
by the possible spectrum of the stochastic background of GWs. 
Given the noise power spectrum of the detector, we compute the SNRs of the GW 
spectrum $\Omega_{\rm  gw}$ using Eq.(\ref{eq:SNR^2_weak}) with 
\begin{equation}
\widetilde{Q}(f) =
\frac{\gamma(|f|)\,\Omega_{\rm filter}(|f|)}{|f|^{3}P_{1}(|f|)
    P_{2}(|f|)}\,.
\label{eq:filter2}
\end{equation}
Here, we consider templates with
a simple power-law type for the optimal filters
\begin{equation}
\Omega_{\rm filter}(f) \propto
{f}^{\alpha}\,,
\label{twostep}
\end{equation} 
where $\alpha$ is a constant. We also consider templates with the change of
the amplitude $f$, that is, $\Omega_{\rm filter}(f) = \Omega_{\rm  gw}(f)$ 
given in Eq. (\ref{real}) for another optimal filter. 
Then, we compare both results and examine whether the template Eq.(\ref{real}) 
is effective not only for the detection of such GWs but also for probing 
the change of the amplitude of their spectrum. 

Before going into the detailed computation, we may have a qualitative 
understanding of SNR for a flat $\Omega_{\rm filter}$.  
{}From Eq.(\ref{eq:filter2}) for a flat $\Omega_{\rm filter}$ and constant $\gamma(f)$, 
a ``V"-shaped noise density $P_i(f)$ becomes a sharper ``$\Lambda$"-shaped optimal filter 
$\widetilde{Q}(f)$ and hence the ``bandwidth''
of $\widetilde{Q}(f)$ is very narrow. 
SNR gets its significant contributions from this bandwidth and does not care about 
the change of the filter beyond that frequency band. So the question is how much 
SNR is improved when $f_c$ falls into the bandwidth of $\widetilde{Q}(f)$. 

\subsection{Sensitivity of LIGO II to stochastic background of GWs}

\begin{figure}[!tbp]
\begin{center}
\includegraphics[width=8cm,angle=0,clip]{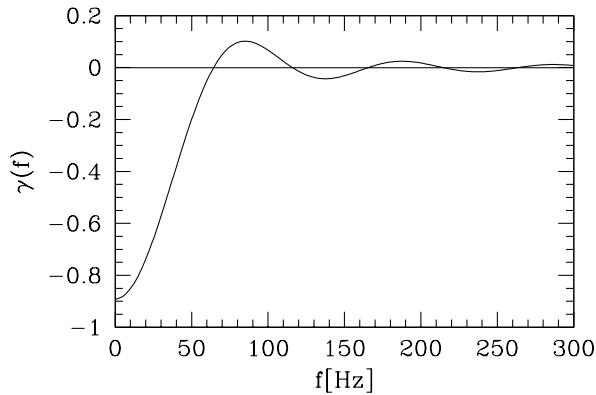}
\end{center}
    \caption{The overlap reduction function $\gamma(f)$ for the Hanford
 and Livingston, LIGO detector pair.}
 \label{fig:overlap}
\end{figure}

\begin{figure}[!tbp]
\begin{center}
\includegraphics[width=8cm,angle=0,clip]{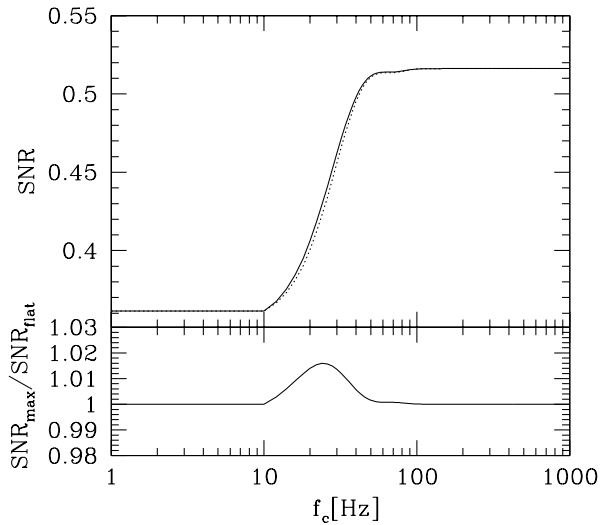}
\end{center}
    \caption{The value (top) and ratio (bottom) of SNRs for 
$\Omega_{\rm filter} = \Omega_{\rm gw}$ (solid line in top panel)
and 
$\Omega_{\rm filter} \propto f^{0}$ (dotted line in top panel)
for LIGO II. We set $d=0.7$.}
 \label{fig:snr_LIGOII}
\end{figure}

First we give the results of SNRs for several templates
in the setup of LIGO II.
For ground based detectors like LIGO II, the sources of noise consist of 
seismic, thermal, and photon shot noises. 
Then, we use the noise power spectrum of the detectors
giving the Figure 1 in \cite{Dalal:2006qt}. The fitting function of this noise
power spectrum is given by \cite{ST}:
\begin{equation}
P_{i}(f)
=  \begin{cases} 
{\mathrm{Max}}[ 10^{-44}(f/10{\rm{Hz}})^{-4} 
+ 10^{-47.25}(f/100{\rm{Hz}})^{-1.7}\,,\,
 	     10^{-46}(f/10^3{\rm{Hz}})^{3} ]\, ; \, 10 < f < 3000 \rm{Hz}\,,
\\
\infty \, ; \, \mathrm{otherwise}\,.
\end{cases}
\label{noise}
\end{equation}
The overlap reduction function $\gamma(f)$ 
is calculated by giving each location for the detector pair.
Fig.~\ref{fig:overlap} shows the overlap reduction function for the Hanford
and Livingston LIGO detector pair \cite{AR}.
Then we compute the SNRs of the GW spectrum $\Omega_{\rm  gw}$ 
using Eq.(\ref{eq:SNR^2_weak}). 
We take 
  $T=10^{7}\sec$ and $\Omega_N= 10^{-10}$.  
The results are shown 
in Figure \ref{fig:snr_LIGOII} and Table \ref{table:LIGO2}. 
We note that SNR is proportional to $\Omega_N$. 

Fig. \ref{fig:snr_LIGOII} shows the dependence of the value and ratio of
${\rm SNR_{max}}$ and ${\rm SNR_{flat}}$ on the critical frequency
$f_{c}$ with damping factor $d=0.7$. 
Here ${\rm SNR_{max}}$ and ${\rm SNR_{flat}}$ are the value of SNR
calculated in Eqs.~(\ref{eq:SNR^2_weak}) and (\ref{eq:filter2}) with
$\Omega_{\rm filter}=\Omega_{\rm gw}$ and 
$\Omega_{\rm filter} \propto f^{0}$, respectively.
We find that $\mathrm{SNR}_{\rm max}$ can differ from $\mathrm{SNR}_{\rm flat}$
around the range $10-50$ Hz which is below the typical frequency range 
of the noise spectrum of LIGO II due to $f^{-3}$ factor in Eq.(\ref{eq:filter2}).
However, the difference is quite small. 
Comparing with the SNR by a flat spectrum, the improvement using the ``true" 
filter $\Omega_{\rm gw}$ is at most 2\% for $d=0.7$. 
In Table \ref{table:LIGO2}, we give results for other $d$ for $f_c=10^{-4}{\rm Hz}$. 
We find no improvement. 
The bandwidth of $\widetilde{Q}(f)$ is very narrow ($10{\rm Hz} \siml f \siml 100{\rm Hz}$).  
SNR gets its significant contributions from this bandwidth and does not care about 
the change of the filter beyond that frequency band. 

We thus conclude that search templates
with a power-law form are sufficient in practice for the detection of GWs
generated during inflation, even though the real spectrum is never
power law.

\begin{table}[tb]
\begin{ruledtabular}
\begin{tabular}{cccccl}
damping rate: $d$ & 0.9 & 0.7 & 0.5 &0.1 & $\Omega_{\rm filter}$  
\\ \hline  \hline  
${\rm SNR}_{\rm max}$ 
& $0.465$ & $0.361$ & $0.258$ & 0.0516 &
$\Omega_{\rm gw}$
\\
${\rm SNR}_{\rm flat}$ & 
$0.465$ & $0.361$ & $0.258$ &0.0516  & power-law with $\alpha =0$  
\\
${\rm SNR}_{\rm PL-1}$ & $0.429$ & $0.333$ & $0.238$ & 0.0476 & 
 power-law with $\alpha =-1$
\\
${\rm SNR}_{\rm PL+1}$& $0.423$ & $0.329$ & $0.235$ & 0.0470 & 
 power law with $\alpha =+1$
\\
\end{tabular}
\end{ruledtabular}
\caption[short]{SNRs for several templates in the setup of LIGO II. 
  We take 
  $T=10^{7}\sec$, $\Omega_N=10^{-9}$, and $f_c=10^{-4}$ Hz. 
  ${\rm SNR}_{\rm max}$ represents SNR for 
  $\Omega_{\rm filter} = \Omega_{\rm gw}$ 
  with $f_c=10^{-4}$ Hz  in Eq.(\ref{real}). 
  ${\rm SNR}_{\rm flat}$ for $\Omega_{\rm filter} \propto f^{0}$, 
  ${\rm SNR}_{\rm BPL \pm 1}$ 
  for $\Omega_{\rm filter} \propto f^{\pm 1}$. Note that SNR is proportional to $\Omega_N$. }
\label{table:LIGO2}
\end{table}

\subsection{Sensitivity of next-generation space interferometers to
  stochastic background of GWs}
\label{subsec:sensitivity}

Finally, we also give the results of SNRs for several templates
with different forms in the setup of next-generation space
interferometers such as FP(Fabry-Perot)-DECIGO \cite{him} and LISA \cite{lisa}.

We consider two detectors forming a starlike
constellation. In the pre-conceptual design, DECIGO is formed by three
drag-free spacecraft, 1000km apart from one another, with observation
frequency band of around $0.1-1.0$Hz. As a result, 
this starlike configuration is identical to the pair of detectors whose
arms are separated by 60 degrees with the separation 
$ 2/\sqrt{3} \times 10^{6} {\mathrm m}$ in the flat ground. 
Because the typical wave length of FP-DECIGO is around 
$10^{8}-10^{9}{\mathrm m}$,
much greater than the separation between detectors, the overlap
reduction function is almost constant in observation
frequency band. Then, we find that $\gamma(f) \sim 0.75$ in the low
frequency limit under the starlike configuration in flat ground.

The signal processing of FP-DECIGO may adopt the same technique as
used in the ground detectors. The essential requirement is that the
relative displacement between the spacecrafts be constant during an
observation.  Adopting the Fabry-Perot configuration, while the
arm-length of the detector can be greatly reduced without changing the
observed frequency range, no flexible combination of time-delayed
signal is possible anymore.  We assume that the output data which is
available for data analysis is only one for each set of detectors.

\begin{figure}[!tbp]
\begin{center}
\includegraphics[width=8cm,angle=0,clip]{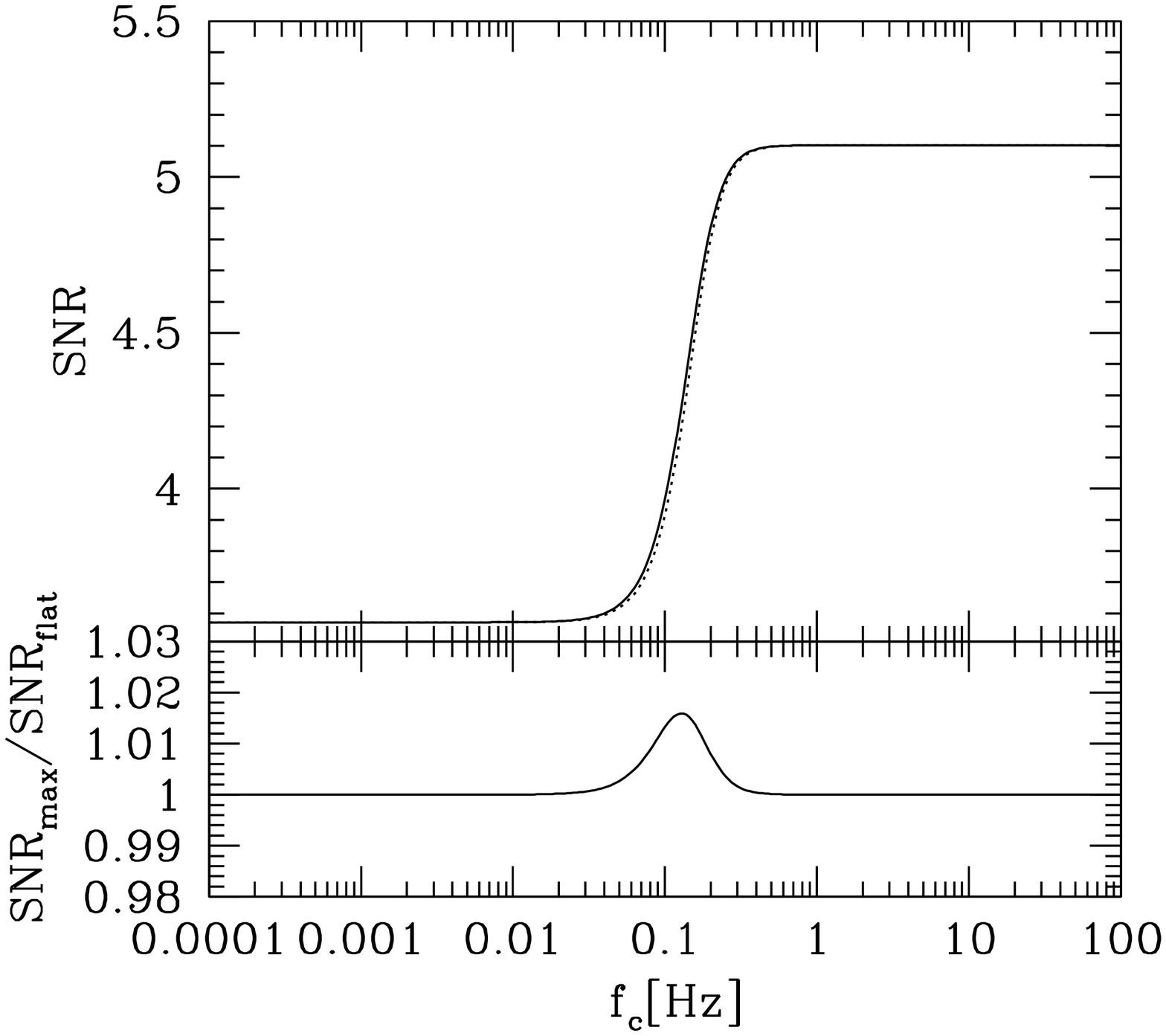}
\end{center}
    \caption{The value (top) and ratio (bottom) of SNRs for 
$\Omega_{\rm filter} = \Omega_{\rm gw}$ (solid line in top panel) and 
$\Omega_{\rm filter} \propto f^{0}$ (dotted line in top panel) for FP-DECIGO. We
 set $d=0.7$.}
 \label{fig:snr_FP}
\end{figure}

\begin{figure}[!tbp]
\begin{center}
\includegraphics[width=8cm,angle=0,clip]{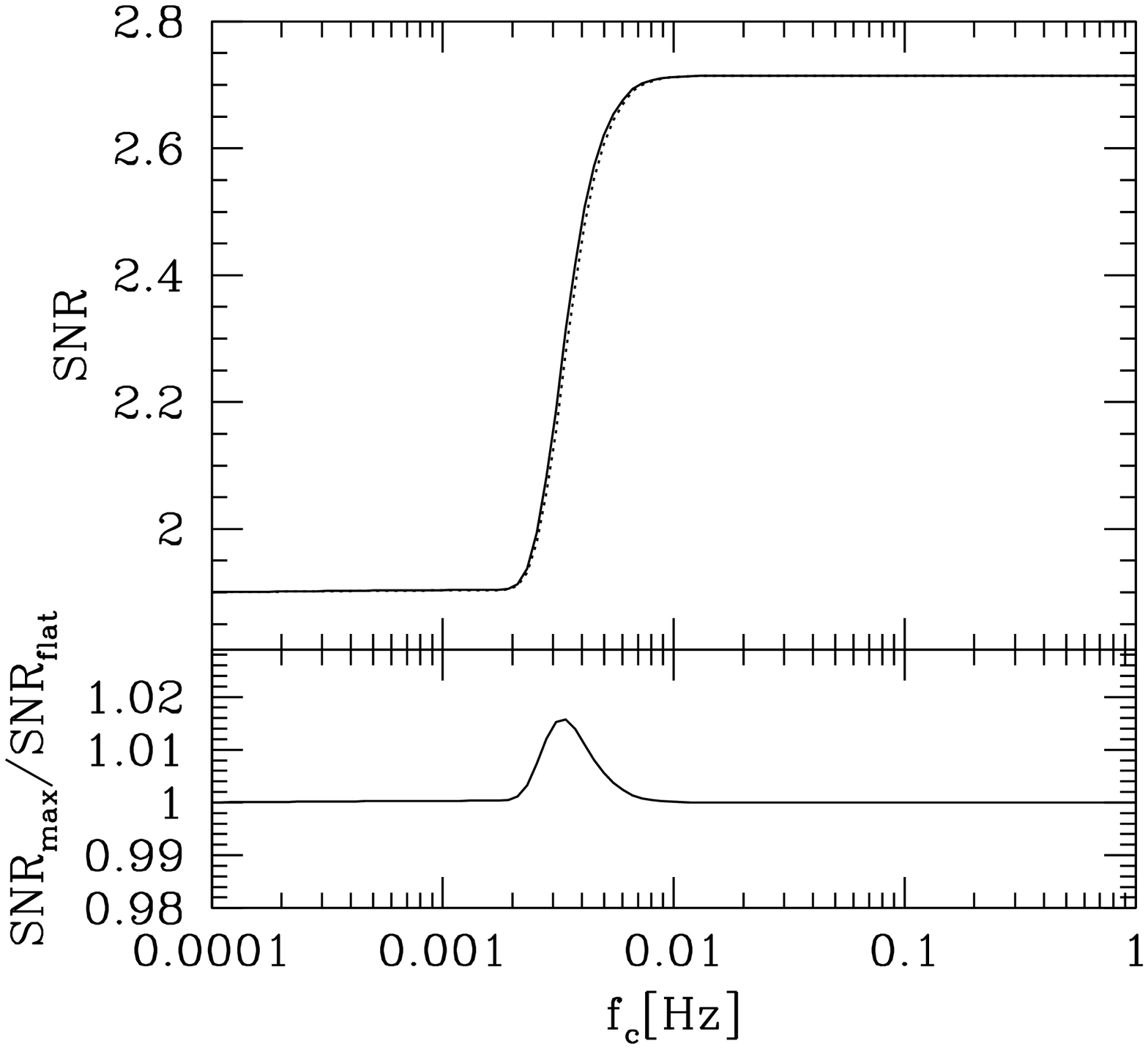}
\end{center}
    \caption{The value (top) and ratio (bottom) of SNRs for 
$\Omega_{\rm filter} = \Omega_{\rm gw}$ (solid line in top panel) and 
$\Omega_{\rm filter} \propto f^{0}$ (dotted line in top panel) for LISA. 
We set $d=0.7$.}
 \label{fig:snr_lisa}
\end{figure}

The sources of noise in FP-DECIGO consist of photon shot noise in the photo detector and 
acceleration noise from the drag-free system and  
radiation pressure noise.  
Each noise spectrum of FP-DECIGO is given in \cite{him}: Photon shot noise is  
$N_{\rm shot}=4.8\times 
10^{-42}(L/{\rm km})^{-2}(1+(f/f_0)^2) ~{\rm Hz}^{-1}$, 
acceleration noise is 
$N_{\rm accl}=4.0\times 10^{-46} (L/{\rm km})^{-2}f^{-4} ~{\rm Hz}^{-1}$, 
radiation pressure noise is  
$N_{\rm rad}=3.6 \times 10^{-51} f^{-4} [{1+(f/f_0)^2}]^{-1} ~ 
{\rm Hz}^{-1} $. 
Here $f_0$ is the characteristic frequency given by $f_0 = 1/4{\mathcal{F}}L$ 
with the fineness of ${\mathcal{F}}=10$ and $L$ is the arm-length and we assume $L=1000$km, 
so that $f_0=7.5$Hz. The noise spectral density is given by the sum: 
$P_{i}(f)=N_{\rm shot}+N_{\rm accl}+N_{\rm rad}$. The detailed discussion of 
noise and instrumental parameters of FP-DECIGO is given in \cite{him}. 
The noise spectrum of LISA (including white-dwarf binaries background \cite{bh}) is taken 
from \cite{lisa-noise}.  
The results for FP-DECIGO are given in Fig.~\ref{fig:snr_FP},
Tables \ref{table:decigo1} and \ref{table:decigo2}. 
We take 
  $T=10^{7}\sec$, $\Omega_N=10^{-15}$ and $f_{0}=7.5$ Hz 
and assume $\gamma (f)=0.75$ as mentioned above. 
The results for LISA are given in Fig.~\ref{fig:snr_lisa} and Table \ref{table:lisa}. 
There we take $T=10^{7}\sec$, 
$\Omega_N=10^{-12}$, $f_c=10^{-4}$Hz and assume $\gamma(f)=0.75$
\footnote{Using more realistic $\gamma(f)$ only reduces the SNR and 
the bandwidth of $\widetilde{Q}(f)$ and does not affect the conclusion. } 
which is maximal for LISA since the angle between the two adjacent laser beams in LISA is $\pi/3$.

Fig. \ref{fig:snr_FP} (Fig. \ref{fig:snr_lisa}) shows the dependence of the value and ratio of
${\rm SNR_{max}}$ and ${\rm SNR_{flat}}$ on the critical frequency
$f_{c}$ with damping factor $d=0.7$ for FP-DECIGO (LISA). 
We find that
both the value and ratio of $\mathrm{SNR}_{\rm max}$ and
$\mathrm{SNR}_{\rm flat}$ increase around $0.1-1 {\rm Hz} (10^{-3}-10^{-2} {\rm Hz})$.
{}From Table \ref{table:decigo1}, we find that the value of SNR in each
filter is independent of the damping factor $d$.
This is because the contribution of the integration for calculation of
SNR  Eq. (\ref{eq:SNR^2_weak})
is the largest around $0.1-1$Hz in all frequency ranges.
On the others hand, Table \ref{table:decigo2} and Table \ref{table:lisa} show that all SNR is
sensitive to the damping factor but is insensitive to the change of the spectrum. 
This is because $\Omega_{\rm gw}$ is regarded
as a flat spectrum with the damping factor $d$ around $0.1-1$Hz ($10^{-3}-10^{-2}$ Hz for LISA)
in the case of $f_c \ll 0.1-1$ Hz ($10^{-3}-10^{-2}$ Hz for LISA). 
As a result, although SNRs themselves 
are enhanced because of much smaller noise power spectrum, 
we find no significant improvement in the use of the filter $\Omega_{\rm gw}$. 

\begin{table}[tb]
\begin{ruledtabular}
\begin{tabular}{cccccl}
damping rate: $d$ & 0.9 & 0.7 & 0.5 &0.1 & $\Omega_{\rm filter}$  
\\ \hline  \hline  
${\rm SNR}_{\rm max}$ 
& $5.10$ & $5.10$ & $5.10$ & 5.10 &
$\Omega_{\rm gw}$
\\
${\rm SNR}_{\rm flat}$ & 
$5.10$ & $5.10$ & $5.10$ &5.10  & $\Omega_{\rm filter}$ with $\alpha =0$  
\\
${\rm SNR}_{\rm PL-1}$ & $4.27$ & $4.27$ & $4.27$ & 4.27 & 
 $\Omega_{\rm filter}$ with $\alpha=-1$
\\
${\rm SNR}_{\rm PL+1}$& $4.60$ & $4.60$ & $4.60$ & 4.60 & 
$\Omega_{\rm filter}$ with $\alpha=+1$
\\
\end{tabular}
\end{ruledtabular}
\caption[short]{SNRs for several templates in the setup of FP-DECIGO. 
  We take 
  $T=10^{7}\sec$, $\Omega_N=10^{-15}$, and $f_{0}=7.5$ Hz. 
  ${\rm SNR}_{\rm max}$ represents SNR for 
  $\Omega_{\rm filter} = \Omega_{\rm gw}$ 
  with $f_c=7.5$ Hz ($=f_0$) in Eq.(\ref{real}). 
  ${\rm SNR}_{\rm flat}$ for $\Omega_{\rm filter} \propto f^{0}$, 
  ${\rm SNR}_{\rm PL \pm 1}$ 
  for $\Omega_{\rm filter} \propto f^{\pm 1}$.}
\label{table:decigo1}
\end{table}

\begin{table}[tb]
\begin{ruledtabular}
\begin{tabular}{cccccl}
damping rate: $d$ & 0.9 & 0.7 & 0.5 &0.1 & $\Omega_{\rm filter}$  
\\ \hline  \hline  
${\rm SNR}_{\rm max}$ 
& $4.59$ & $3.57$ & $2.55$ & $0.510$ &
$\Omega_{\rm gw}$
\\
${\rm SNR}_{\rm flat}$ & 
$4.59$ & $3.57$ & $2.55$ & $0.510$ & $\Omega_{\rm filter}$ with $\alpha =0$  
\\
${\rm SNR}_{\rm PL-1}$ & $3.75$ & $2.92$ & $2.08$ & $0.417$ & 
 $\Omega_{\rm filter}$ with $\alpha =-1$
\\
${\rm SNR}_{\rm PL+1}$& $4.14$ & $3.22$ & $2.30$ & $0.460$ & 
$\Omega_{\rm filter}$ with $\alpha =+1$
\\
\end{tabular}
\end{ruledtabular}
\caption[short]{The same as Table \ref{table:decigo1} except $f_c=10^{-4}{\rm Hz}$ 
in Eq.(\ref{real}).}
\label{table:decigo2}
\end{table}

\begin{table}[tb]
\begin{ruledtabular}
\begin{tabular}{cccccl}
damping rate: $d$ & 0.9 & 0.7 & 0.5 &0.1 & $\Omega_{\rm filter}$  
\\ \hline  \hline  
${\rm SNR}_{\rm max}$ 
& $2.44$ & $1.90$ & $1.36$ & $0.276$ &
$\Omega_{\rm gw}$
\\
${\rm SNR}_{\rm flat}$ & 
$2.44$ & $1.90$ & $1.36$ &$0.272$  & $\Omega_{\rm filter}$ with $\alpha =0$  
\\
${\rm SNR}_{\rm PL-1}$ & 
$1.46$ & $1.40$ & $0.822$ & $0.187$ & 
 $\Omega_{\rm filter}$ with $\alpha=-1$
\\
${\rm SNR}_{\rm PL+1}$& 
$2.32$ & $1.80$ & $1.29$ & $0.258$ & 
$\Omega_{\rm filter}$ with $\alpha=+1$
\\
\end{tabular}
\end{ruledtabular}
\caption[short]{SNRs for several templates for LISA. 
  We take   $T=10^{7}\sec$, $\Omega_N=10^{-12}$ and $f_c=10^{-4}$ Hz. }
\label{table:lisa}
\end{table}

\section{Discussions and Summary}

Motivated by the fact that the spectrum of gravitational waves 
has fine structures due to the change 
of the number of relativistic degrees 
of freedom, we have investigated the 
validity of the use of the templates of 
power-law shape as search templates 
for a stochastic gravitational-wave background. 
Comparing the SNR using the template of power-law shape and 
the SNR using the template of step function shape, 
we find that the resulting SNR is insensitive 
to the change of the amplitude. 

Although we have focused on the change of the spectrum associated with 
electroweak phase transition, our analysis is not limited to 
it and is easily extended to other frequency ranges. 

Our results have both bad news and good news. 
The bad news is that gravitational wave measurements with the amplitude
close to their detection threshold are insensitive to 
the detailed structure of the spectrum. Although the spectrum beyond 
$10^{-3}$Hz depends on the particle physics 
beyond 1 TeV, gravitational waves 
measurements themselves do not discern the 
difference: probing SUSY via gravitational- 
wave observations is not feasible if its amplitude is so small that a long time 
observation is required to achieve sufficient SNR as discussed here.

The spectrum of a stochastic gravitational wave background may not be determined by 
single observations. In order to determine
 the spectrum (in particular, the spectral 
index) of gravitational waves, multiple 
observations at different frequencies are 
required. If such observations are realized in the future, the change of
the number of relativistic degrees of freedom may be detected assuming
that the power of the spectrum is known independently.

The bad news becomes good news at the same time.  
That is, the templates of simple power-law shape are sufficient 
as  search templates for stochastic gravitational 
waves: no detailed transfer function, 
 which is dependent on particle physics models for $f>10^{-3}$Hz, 
is necessary.

\section*{Acknowledgments}

We would like to thank Atsushi Taruya for careful reading of the manuscript 
and for providing us useful comments. 
This work was partially supported by JSPS Grant-in-Aid for Scientific
Research No.~16340076, No.~19340054(JY), No.~17204018(TC), 
and No.~18740157(MY).  Also,
T.~C.\/ is supported in part by Nihon University. M.Y. is supported in
part by the project of the Research Institute of Aoyama Gakuin
University.

\end{document}